\documentclass[aps,prl,twocolumn,superscriptaddress,showpacs]{revtex4-1}
\usepackage[colorlinks=true,allcolors=blue]{hyperref}
\usepackage{graphicx,braket,amsmath}

\begin{document}

\renewcommand*{\figureautorefname}{Fig.}

\title{Single-shot readout and relaxation of singlet/triplet states in exchange-coupled $^{31}$P electron spins in silicon}

\author{Juan P. Dehollain}
\author{Juha T. Muhonen}
\author{Kuan Y. Tan}
\altaffiliation{Present address: QCD Labs, COMP Centre of Excellence, Department of Applied Physics, Aalto University, 00076 Aalto, Finland}
\affiliation{Centre for Quantum Computation and Communication Technology}
\affiliation{School of Electrical Engineering and Telecommunications, UNSW Australia, Sydney NSW 2052, Australia}
\author{Andre Saraiva}
\affiliation{Instituto de Fisica, Universidade Federal do Rio de Janeiro, Caixa Postal 68528, 21941-972 Rio de Janeiro, Brazil}
\affiliation{University of Wisconsin-Madison, Madison, Wisconsin 53706, USA}
\author{David N. Jamieson}
\affiliation{Centre for Quantum Computation and Communication Technology}
\affiliation{School of Physics, University of Melbourne, Melbourne, VIC 3010, Australia}
\author{Andrew S. Dzurak}
\author{Andrea Morello}
\email{a.morello@unsw.edu.au}
\affiliation{Centre for Quantum Computation and Communication Technology}
\affiliation{School of Electrical Engineering and Telecommunications, UNSW Australia, Sydney NSW 2052, Australia}

\date{\today}

\begin{abstract}
We present the experimental observation of a large exchange coupling $J \approx 300$~$\mu$eV between two $^{31}$P electron spin qubits in silicon. The singlet and triplet states of the coupled spins are monitored in real time by a Single-Electron Transistor, which detects ionization from tunnel-rate-dependent processes in the coupled spin system, yielding single-shot readout fidelities above 95\%. The triplet to singlet relaxation time $T_1 \approx 4$~ms at zero magnetic field agrees with the theoretical prediction for $J$-coupled $^{31}$P dimers in silicon. The time evolution of the 2-electron state populations gives further insight into the valley-orbit eigenstates of the donor dimer, valley selection rules and relaxation rates, and the role of hyperfine interactions. These results pave the way to the realization of 2-qubit quantum logic gates with spins in silicon, and highlight the necessity to adopt gating schemes compatible with weak $J$-coupling strengths.
\end{abstract}


\maketitle

Entangling two-qubit operations, together with single-qubit rotations, form a universal set of quantum logic gates for circuit-based quantum computing~\cite{Bennett2000n}. These have been demonstrated in several physical qubit platforms~\cite{Ladd2010n}, including spins in semiconductors~\cite{Petta2005s,shulman12S,Nowack2011s}. The best qubit coherence times in the solid state have been obtained with spins in isotopically purified silicon~\cite{Tyryshkin2012nm,Steger2012s,Muhonen2014x} and carbon~\cite{Balasubramanian2009nm,Maurer2012s}. Additionally, silicon is the material that underpins all of modern electronics, which makes it an appealing candidate for spin-based quantum technologies~\cite{kane1998n,Hollenberg2006prb,zwanenburg13rmp}. This platform can be scaled to the single-atom limit by using industry-compatible ion-implantation~\cite{Jamieson2005apl} or atomically-precise scanning tunneling microscopy~\cite{Fuechsle2012nn} fabrication methods. The coherent operation of spin-based qubits in Si has been demonstrated in single $^{31}$P donor atoms~\cite{Pla2012n,Pla2013n,Muhonen2014x} and double quantum dots~\cite{Maune2012n,kim14CM}. Conversely, an entangling quantum logic gate for a pair of spin qubits in silicon is still awaiting experimental demonstration. Several coupling mechanisms can be used for this purpose, including magnets~\cite{Trifunovic2013prx} and microwave photons~\cite{Hu2012prb}, but the simplest is the exchange interaction $J$, arising from the overlap of electron wavefunctions~\cite{Koiller2002prl,Wellard2003prb}.

Exchange interaction between pairs of donors in silicon has been observed in bulk spin resonance experiments~\cite{Jerome1964pr} and, very recently, by electron transport experiments through a donor molecule~\cite{gonzalez13CM}. However, relevant applications to quantum information processing require the ability to measure the instantaneous quantum state of the qubits. Here we report the observation of large exchange coupling $J \approx 300$~$\mu$eV between the electrons of a $^{31}$P donor pair. Additionally, the $^{31}$P pair is integrated within a top-gated silicon Single-Electron Transistor (SET)~\cite{Angus2007nl} to perform single-shot readout of the spin singlet ($\ket{S}=\left(\ket{\uparrow \downarrow}-\ket{\downarrow \uparrow}\right)/\sqrt{2}$) and triplet ($\ket{T_-}=\ket{\downarrow \downarrow}, \ket{T_0}=\left(\ket{\uparrow \downarrow}+\ket{\downarrow \uparrow}\right)/\sqrt{2}, \ket{T_+} = \ket{\uparrow \uparrow}$) states of the two-electron system. We exploit the significant difference in the size of the orbital wavefunctions for $\ket{S}$ and $\ket{T}$ states to demonstrate high-fidelity tunnel-rate-selective readout (TR-RO)~\cite{Hanson2005prl}. We apply these techniques to measure the valley and spin relaxation times, and their dependence on the external magnetic field $B$.

\begin{figure}[b]
 \includegraphics[width=\columnwidth]{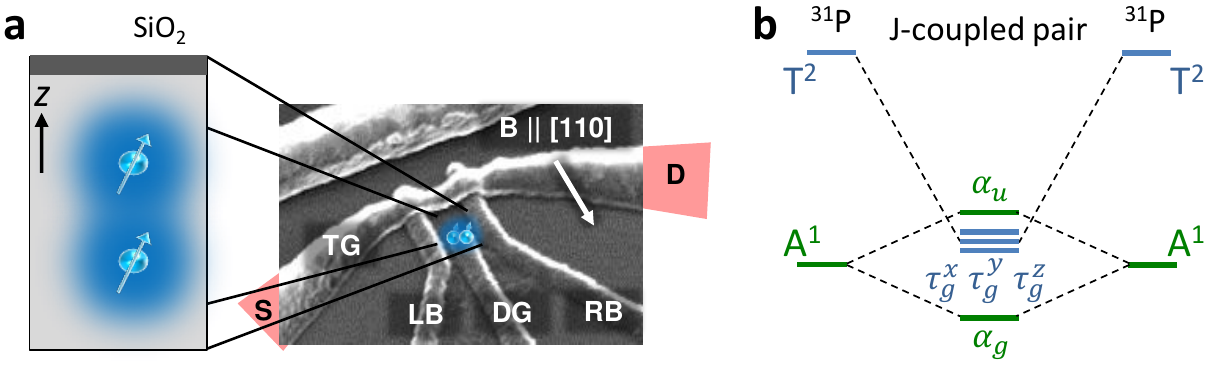}
 \caption{\textbf{a.} Scanning electron microscope image of a device similar to the one used in the experiments. The gates TG, LB, RB along with the S, D diffusion regions make up the Single Electron Transistor. A static magnetic field $B$ can be applied in the plane of the device, along the [110] Si crystal axis. Inset: sketch of the two $^{31}$P donors, aligned along the $z$ axis. \textbf{b.} Diagram showing the expected modification of the valley-orbit states for coupled $^{31}$P donors $\lesssim 6$~nm apart.\label{fig:device}}
\end{figure}

The device was obtained from the same batch as the one described in Ref.~\cite{Pla2012n}. It consists of a natural silicon substrate implanted with phosphorus ions~\cite{Jamieson2005apl}. The donor electrons are tunnel-coupled to the island of a top-gated SET~\cite{Morello2009prb} (\autoref{fig:device}a). The electrochemical potential of the donor electrons $\mu_D$ can be varied using a donor gate (DG) above the implant window.

\begin{figure}
 \includegraphics[width=\columnwidth]{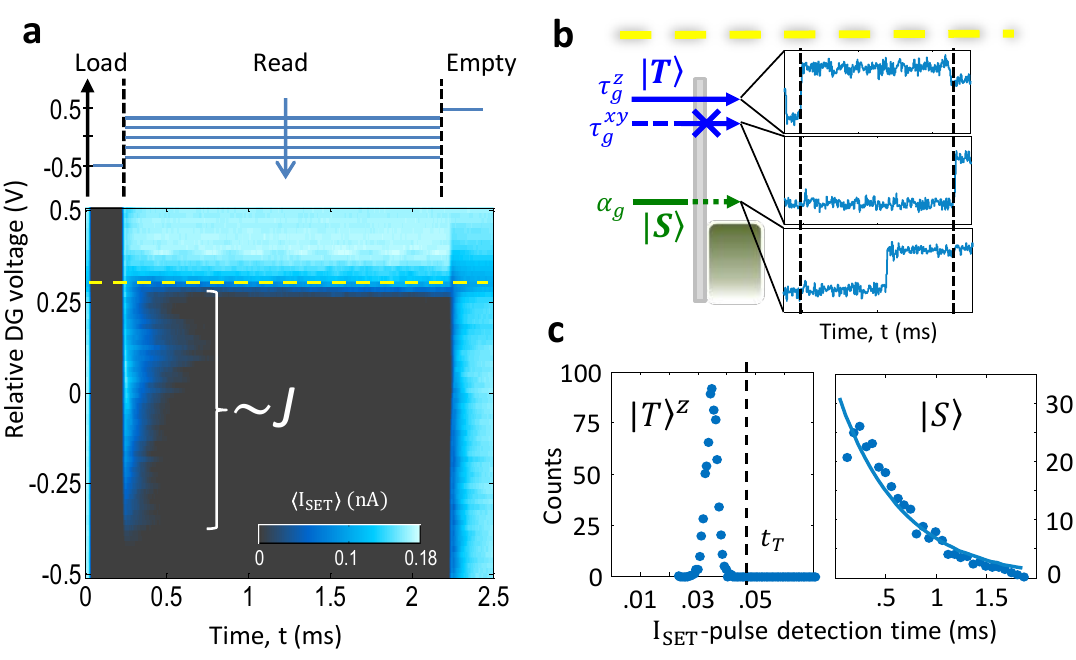}
 \caption{\textbf{a.} Three-phase pulse sequence and averaged SET current $\langle I_{\rm SET} \rangle$, used to estimate the exchange coupling $J$, at $B=0$. The dashed line identifies the appropriate read-phase voltage for tunnel-rate selective (TR-RO) readout. \textbf{b.} Diagrams of the electrochemical potentials $\mu_S, \mu_{T^{x,y}}, \mu_{T^z}$ relative to the SET Fermi energy $E_F$ with examples of readout traces identifying each of the states. Due to the valley configuration of the SET island, $\ket{T}$ is only allowed to tunnel if it occupies the $\tau_g^z$ state. \textbf{c.} TR-RO fidelity: histograms of the detection times of a pulse in $I_{\rm SET}$ during the read-phase, used to extract the readout fidelities~\cite{supp}. \label{fig:readout}}
\end{figure}

With the DG voltage ($V_{\rm DG}$) set near a donor charge transition, the device is tuned so the SET current switches between $I_{\rm SET}=0$ (Coulomb-blockade) and $I_{\rm SET}\neq0$ when the system is neutral or ionized, respectively. We use a 3-level single-shot spin readout sequence~\cite{Elzerman2004n} (\autoref{fig:readout}a) consisting of load, read and empty phases. During the read-phase, we measure $I_{\rm SET}$ while varying $V_{\rm DG}$ such that the $\mu_D$ goes from higher to lower than $E_F$. A well-defined ``tail'' (\autoref{fig:readout}a) where excess current occurs at the start of the read-phase indicates the presence of an energy-split pair of electron states. The high-energy electron tunnels out of the donor ($I_{\rm SET}\neq0$) shortly after the start of the read-phase, and is replaced by one in the low energy state ($I_{\rm SET}=0$ again) thereafter. The data in \autoref{fig:readout} was taken in the absence of magnetic field ($B=0$~T). Therefore the observed splitting cannot be the Zeeman energy $E_z = h\gamma_e B$ ($\gamma_e \approx 28$~GHz/T is the electron gyromagnetic ratio) of a single spin~\cite{Morello2010n}. We postulate that the measurement in \autoref{fig:readout}a constitutes the observation of the $\ket{S}$ and $\ket{T}$ states of a pair of $^{31}$P donors, split by an exchange interaction $J = \mu_T - \mu_S$, where $\mu_T$ and $\mu_S$ are the $\ket{T}$ and $\ket{S}$ electrochemical potentials at $B=0$. To extract the value of $J$ we first convert $V_{\rm DG}$ to a shift in $\mu$, by fitting a Fermi distribution function to the shape of $I_{\rm SET}(V_{\rm DG})$ for $0.25 < V_{\rm DG} < 0.35$~V in the read-phase after the decay of the ``tail'', and using the value $T_{el} = 125 \pm 25$~mK (measured separately) to calibrate the energy scale. Then, the length of the readout ``tail'' $\Delta V_{\rm DG} = 0.6 \pm 0.1$~V can be converted into the value of $J = 345 \pm 100$~$\mu$eV. This value of $J$ is expected to correspond to donors $< 8$~nm apart~\cite{Koiller2002prl,Wellard2003prb,gonzalez13CM}.

Tuning the device to the region indicated by the dashed line in \autoref{fig:readout}a, the single-shot readout traces reveal two distinct tunnel-out processes (shown in \autoref{fig:readout}c): A slow process with a tunnel time $\approx 0.9$~ms, and a faster process for which the tunnel time is shorter than the rise-time $\approx 35$~$\mu$s of the amplifier (see \autoref{fig:readout}b for sample traces). The observation of two very distinct tunnel rates reinforces the interpretation that we are observing the spin states of a $J$-coupled donor pair. The $\ket{T}$ state must correspond to an excited 2-electron orbital, with a more extended wavefunction~\cite{Kouwenhoven2001rpp} that results in stronger tunnel coupling to the nearby SET island.

The \{1s\} orbital of a single $^{31}$P donor in Si has a valley-orbit ground state $A^1$ (1-fold degenerate), and excited states $T^2$ (3-fold degenerate) and $E$ (2-fold degenerate)~\cite{Kohn1955pr}. In particular, the 3-fold degeneracy of $T^2$ arises from it being an antisymmetric combination of pairs of valleys $\pm x, \pm y, \pm z$, where all valleys have the same energy. The $A^1$ to $T^2$ splitting is $\approx 11.7$~meV, making the excited valley-orbit states unimportant for most aspects of single-qubit physics. However, in a donor pair with strong exchange interaction, the hybridization of the valley-orbit states results in ``bonding'' / ``antibonding'' eigenstates, whose energy is split according to the wavefunction overlap. The Bohr radius of the $T^2$ states is about twice that of $A^1$, resulting in a much larger splitting of the coupled states. It has been estimated~\cite{Klymenko2014jp} that for interdonor separation $\lesssim 6$~nm there is an inversion in the hierarchy of states that originate from single-donor $A^1$ and $T^2$. The energy of the bonding combination of $T^2$ states ($\tau_g$) crosses below that of the antibonding $A^1$ ($\alpha_u$), whereas the overall ground state always remains the bonding $A^1$ combination ($\alpha_g$) (see \autoref{fig:device}b). Therefore, in this configuration the spin-singlet state occupies the $\alpha_g$ valley-orbit eigenstate, while the spin-triplets can occupy any of the three $\tau_g^{x,y,z}$ states, distinguished by their valley composition. We denote all the available triplet states as $\ket{T_{+,0,-}}^{x,y,z} = |\alpha_g \tau_g^{x,y,z}| \otimes \ket{T_{+,0,-}}$, where  $| ... |$ stands for the Slater determinant.

Two crucial aspects of the physics of donors and dots in silicon need to be considered here. First, the 2-electron $\tau_g^{x,y,z}$ states are not degenerate. Consider for example a donor pair oriented along $z$, as in \autoref{fig:device}a.  Since the transverse effective mass in Si is smaller than the longitudinal one \cite{Ando1982rmp}, states composed of valleys perpendicular to the orientation of the pair have stronger tunnel coupling, hence $\tau_g^{x,y}$ are lowered in energy further than the $\tau_g^z$ state (\autoref{fig:device}b). Similarly, $\alpha_g$ is not an equal-weight combination of all 6 valleys, but has a predominant component of valleys perpendicular to the dimer axis. Second, the spin state of the donor pair is read out through electron tunneling into the island of an SET formed at a [001] interface, where the electron states consist exclusively of $\pm z$ valleys. As a consequence, the SET island only couples to states of the donor dimer with nonzero $\pm z$ valley composition. Both these Si-specific aspects are revealed in the time-resolved experiments described below.

Single-shot TR-RO~\cite{Hanson2005prl} is performed by setting a time threshold $t_T$, a maximum readout wait time $t_R = 2$~ms, and declaring that (i) a tunnel-out event detected at $t < t_T$ corresponds to $\ket{T}^z$, (ii) $t_T < t < t_R$ corresponds to $\ket{S}$, and (iii) no tunnel event within $t_R$ corresponds to $\ket{T}^{x,y}$ states that do not couple to the SET. A statistical analysis~\cite{supp} of the histograms in \autoref{fig:readout}c using these thresholds reveals a TR-RO readout fidelity $\approx 95$\% for $\ket{T}^z$--$\ket{S}$ discrimination and $\approx 90$\% for $\ket{S}$--$\ket{T}^{x,y}$ discrimination.

\begin{figure}
 \includegraphics[width=\columnwidth]{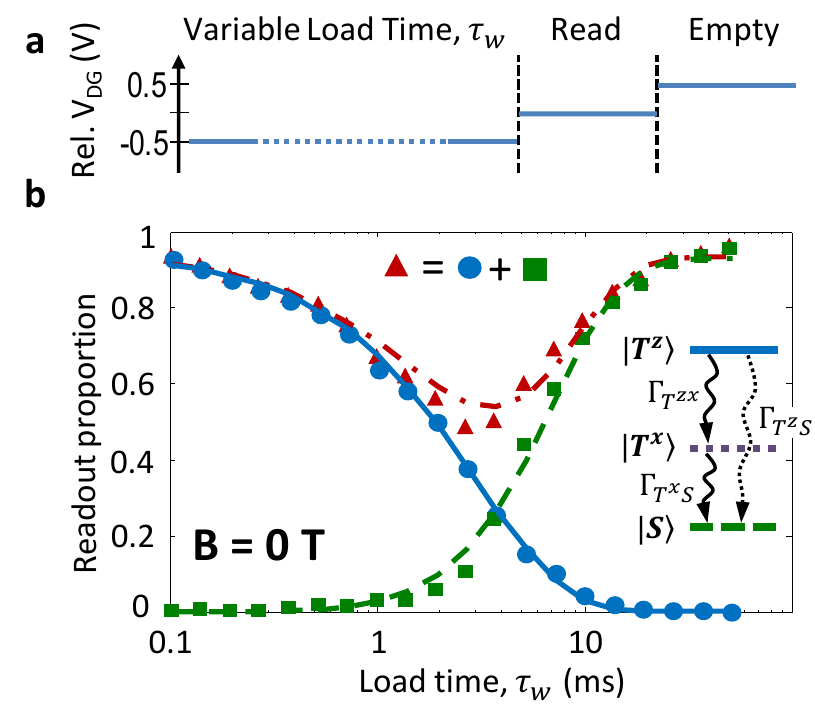}
 \caption{\textbf{a.} Gate pulsing scheme for relaxation measurements. \textbf{b.} Spin and valley relaxation at $B=0$. Dots: fraction of fast tunneling events identified as $\ket{T}^z$ according to the TR-RO threshold. Squares: slow tunneling events, identified as $\ket{S}$, excluding those with no observed tunneling in the read-phase ($\ket{T}^x$). Solid lines are fits to the model in \autoref{model:B0}. The triangles correspond to $T^z + S$. \label{fig:b0}}
\end{figure}

\begin{figure}
 \includegraphics[width=\columnwidth]{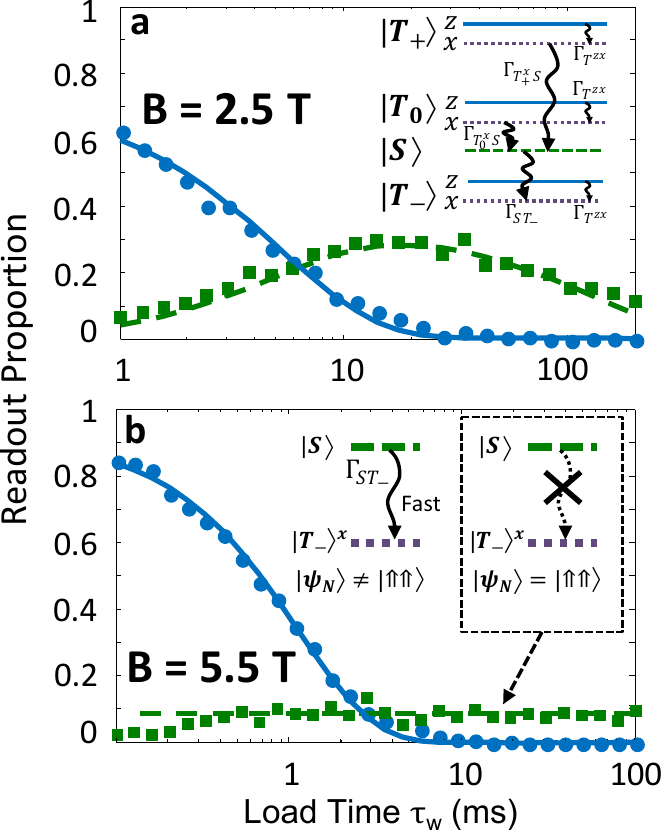}
 \caption{Relaxation in different $B$-field regimes. \textbf{a.} $B=2.5$~T, where $E_Z \gtrsim J$ and $\Gamma_{ST^x_-}$ can be observed. Solid lines are fits to \autoref{model:Bn0}. Inset shows the spin and valley energies with all the available relaxation channels. \textbf{b.} At $B=5.5$~T, $E_Z \gg J$ and $\Gamma_{ST^x_-}$ becomes too fast to resolve. Solid line: fit to \autoref{model:Bn0}. The long-time plateau of $S$ is due to nuclear spin selection rules sketched in the inset (see text). \label{fig:bn0}}
\end{figure}

This readout technique allows us to follow in real time the evolution of the state populations, as they relax from the initially loaded state to the ground state. We do this by taking repeated readout traces as a function of the duration $\tau_w$ of the load phase, and calculating the readout proportion of $\ket{T}^z$ (dots in Figures \ref{fig:b0} and \ref{fig:bn0}) or $\ket{S}$ (squares). The result of this measurement at $B=0$ is plotted in \autoref{fig:b0}. The sum of the $\ket{S}$ and $\ket{T}^z$ detection probabilities as a function of load time $\tau_w$ is not constant, but exhibits a dip for $\tau_w \approx 1 - 10$~ms. This can be explained by assuming we have a donor dimer oriented (predominantly) along $z$, such that the $\ket{T}^{x,y}$ are below $\ket{T}^z$ in energy, acting as ``shelving'' states in the relaxation process. We model the data with the rate equations below, where we include for simplicity only the $\ket{T}^x$ shelving state:
\begin{align}
dT^z/d\tau_w &= - (\Gamma_{T^{zx}} + \Gamma_{T^zS})T^z \nonumber \\
dT^x/d\tau_w &= \Gamma_{T^{zx}}T^z - \Gamma_{T^xS}T^x \label{model:B0} \\
dS/d\tau_w &= \Gamma_{T^zS}T^z + \Gamma_{T^xS}T^x \nonumber
\end{align}

Here $T^z$, $T^x$, $S$ are the populations of the corresponding states, $\Gamma_{T^{zx}}$ is the relaxation rate from $\ket{T}^z$ to $\ket{T}^x$ and $\Gamma_{T^{z(x)}S}$ is the $\ket{T}^{z(x)}$ to $\ket{S}$ relaxation rate. We include the parameters $c_{T} \equiv T^z|_{\tau_w=0}$ and $c_{S} \equiv S|_{\tau_w = \infty}$ ($\in [0,1]$) that multiply the corresponding populations to account for initialization and measurement imperfections. A least-squares fit to the data in \autoref{fig:b0}b yields $\Gamma_{T^{zx}}^{-1}=2.9 \pm 0.2$~ms, $\Gamma_{T^xS}^{-1}=4.1 \pm 0.4$~ms, $c_{T} = 0.94 \pm 0.02$, and $c_{S} = 0.93 \pm 0.02$. The model also yields $\Gamma_{T^zS}\ll\Gamma_{T^xS}$ (an accurate value of $\Gamma_{T^zS}$ could not be extracted). This is again consistent with having a dimer along $z$, for which it is predicted that -- in the high-$J$ regime -- the valley composition of the ground state $\alpha_g$ will have five times less contribution from the valleys which are longitudinal to the dimer orientation~\cite{Klymenko2014jp}. Finally, the near-unity value of $c_{T^z}$ confirms that the system is preferentially initialized in $\ket{T}^z$, as expected on the basis of the spatial extent of $\tau_g^z$ states, and the valley selection rules discussed above.

In this picture, $\Gamma_{T^{zx}}$ represents a valley relaxation rate, while the spin relaxation process is captured by $\Gamma_{T^{x}S}$. The value of $\Gamma_{T^{x}S}^{-1} \equiv T_1 \approx 4$~ms extracted from the data agrees well with the $\ket{T} \rightarrow \ket{S}$ relaxation times predicted by Borhani and Hu~\cite{Borhani2010prb} specifically for $^{31}$P donor pairs in Si, in the presence of an exchange interaction $J \approx 300$~$\mu$eV. The electron-nuclear hyperfine coupling $A$ (here $\ll J$) mixes the $J$-split $\ket{S},\ket{T}$ states and provides a new channel for spin-lattice relaxation which is $\approx 3$~orders of magnitude faster than a single-spin flip at an equivalent value of the Zeeman splitting ($E_Z \approx 300$~$\mu$eV corresponds to $B\approx 2.5$~T on a single spin, where $T_1 \approx 1$~s~\cite{Morello2010n}). The $\ket{T} \rightarrow \ket{S}$ relaxation is predicted to slow down at lower $J$, giving $T_1 \gg 1$~s for $J \approx 1$~$\mu$eV. For $J < A = 117$~MHz~$\approx 0.5$~$\mu$eV this relaxation channel becomes suppressed. Therefore our measurements clearly indicate that 2-qubit coupling schemes which do not require large values of $J$~\cite{kalra13CM,Srinivasa2013x} will have the additional benefit of preserving the long spin lifetime of the individual qubits.

Applying a magnetic field $B \parallel [110]$ splits the $\ket{T}$ states by $E_Z$. For $J \approx 300$~$\mu$eV, $E_Z < J$ when $B \lesssim 2.5$~T. In this regime, we found no $B$-dependence of $\Gamma_{T^{zx}}$ (data not shown), as expected for orbital relaxation at low fields~\cite{Murdin2013nc}. The data was not conclusive enough to extract further information on the spin relaxation rate $\Gamma_{T^{z}S}$. At $B=2.5$~T, $E_Z \gtrsim J$ (\autoref{fig:bn0}a), $\ket{T_-}$ becomes the spin ground state. Assuming that the Zeeman-split $\ket{T}^z$ states load with equal probability, and neglecting the single-spin relaxation channels between triplet states (for which $\Gamma^{-1} \approx 1$~s at 2.5 T~\cite{Morello2010n}), the rate equation model becomes:
\begin{align}
dT^z_+/d\tau_w &= -\Gamma_{T^{zx}}T^z_+ \nonumber \\
dT^x_+/d\tau_w &= \Gamma_{T^{zx}}T^z_+ - \Gamma_{T^x_+S}T^x_+ \nonumber \\
dT^z_0/d\tau_w &= -\Gamma_{T^{zx}}T^z_0 \nonumber \\
dT^x_0/d\tau_w &= \Gamma_{T^{zx}}T^z_0 - \Gamma_{T^x_0S}T^x_0 \label{model:Bn0} \\
dS/d\tau_w &= \Gamma_{T^x_+S}T^x_+ + \Gamma_{T^x_0S}T^x_0 - \Gamma_{ST^x_-}S \nonumber \\
dT^z_-/d\tau_w &= -\Gamma_{T^{zx}}T^z_- \nonumber \\
dT^x_-/d\tau_w &= \Gamma_{ST^x_-}S + \Gamma_{T^{zx}}T^z_- \nonumber
\end{align}
 A fit to the data in \autoref{fig:bn0}a yields $\Gamma_{T^{zx}}^{-1}= 5.2\pm0.4 $~ms, $\Gamma_{ST^x_-}^{-1}= 146\pm25 $~ms, $\Gamma_{T^x_+S}^{-1},\Gamma_{T^x_0S}^{-1} \ll \Gamma_{ST^x_-}^{-1}$, $c_{T}= 0.72\pm0.04 $, and $c_{S}= 0.50\pm0.03 $. Since $\Gamma_{T^{zx}}^{-1} \ll \Gamma_{ST^x_-}^{-1}$, the $\ket{S}$ population as a function of $\tau_w$ first increases ($\ket{T_{+,0}}{\rightarrow}\ket{S}$) then decreases ($\ket{S}{\rightarrow}\ket{T_-}$).

When $B \gtrsim 4$~T, $\Gamma_{ST^x_-}$ becomes the fastest rate, and at $B=5.5$~T (\autoref{fig:bn0}b) only $\Gamma_{T^{zx}}^{-1}= 1.13\pm0.13 $~ms, with $c_{T}= 0.92\pm0.05 $, can be reliably extracted from the data. Interestingly, we observe a constant population of $\ket{S}$ for $\tau_w \gtrsim 1$~ms. This reveals a subtle feature of the spin relaxation mechanism of Borhani and Hu~\cite{Borhani2010prb}: the hyperfine interaction $A$ mixes states having the same total value of the electron ($m_e$) and nuclear ($m_N$) spin quantum number. The transition $\ket{S}{\rightarrow}\ket{T_-}$ yields $\Delta m_e = -1$, thus requires $\Delta m_N = +1$, and becomes forbidden if the $^{31}$P nuclei are in the state $\ket{\psi_N} = \ket{\Uparrow \Uparrow}$. We interpret the long-time plateau of $S$ as a manifestation of this spins selection rule. The plateau height should depend on the probability that $\ket{\psi_N} =\ket{\Uparrow \Uparrow}$, which is unknown and uncontrolled in this experiment, but we may assume that the nuclei randomly populate all possible states over the time necessary to acquire a set of data as in \autoref{fig:bn0}.

The time-resolved observation of singlet and triplet states of an exchange-coupled $^{31}$P donor pair reported here provides a physical basis for the construction of large-scale donor-based quantum computer architectures~\cite{Hollenberg2006prb}. The short $\ket{T}{\leftrightarrow}\ket{S}$ relaxation times $T_1 \approx 4$~ms in this experiment arise from the interplay of a large exchange coupling $J \approx 300$~$\mu$eV with the hyperfine interaction $A = 117$~MHz~$\approx 0.5$~$\mu$eV. Therefore, our results indicate that the best regime to operate $J$-mediated 2-qubit logic gates is where $J \lesssim A$, as described in recent proposals~\cite{kalra13CM,Srinivasa2013x}.

\begin{acknowledgments}
We acknowledge discussions with M. House, T. Watson and X. Hu. This research was funded by the Australian Research Council Centre of Excellence for Quantum Computation and Communication Technology (project number CE110001027) and the US Army Research Office (W911NF-13-1-0024). We acknowledge support from the Australian National Fabrication Facility. A.S. acknowledges the William F. Vilas Trust for financial support.
\end{acknowledgments}

%%% RERUN IF THERE ARE NEW REFERENCES %%%
%\bibliography{QC_st}

%

\clearpage
\onecolumngrid

\section*{Supplemental material for ``Single-shot readout and relaxation of singlet/triplet states in exchange-coupled $^{31}$P electron spins in silicon"}
\setcounter{figure}{0}
\renewcommand{\thefigure}{S\arabic{figure}}
\setcounter{page}{1}
\renewcommand{\thepage}{S\arabic{page}}
\renewcommand*{\theHfigure}{\thepart.\thefigure}

\subsection{Readout fidelities}

\begin{figure}[b]
 \includegraphics[width=\textwidth]{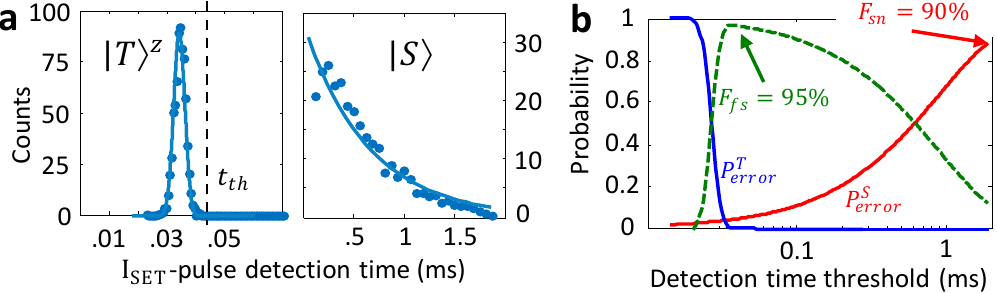}
 \caption{\textbf{a} Histograms of tunnel-out events taken from a data-set of 1000 readout traces. Both histograms are constructed from the same data-set, but use different bin resolution to highlight different tunneling processes. The histogram on the left shows the peak arising from fast tunneling events detected with limited bandwidth, fitted with a normal distribution with $\mu_d = 27$~$\mu$s and $\sigma_d = 3.2$~$\mu$s. The histogram on the right shows an exponential decay with $\Gamma_{S,\mathrm{out}} = 1.1 \times 10^3$~s$^{-1}$ dictated by the $\ket{S}$ tunnel-out time. The first point of the histogram is omitted to exclude the fast-tunnel events. \textbf{b} Measurement fidelities for discriminating between fast ($\ket{T}^z$) and slow ($\ket{S}$) tunneling events ($F_{\rm fs}$, green dashed line), and between slow $\ket{S}$) and no ($\ket{T}^{xy}$) tunneling ($F_{\rm sn}$, red line), extracted from fits to the histograms.\label{fig:fid}}
\end{figure}

We can assess the fidelity of the tunnel-rate-selective readout technique by constructing a histogram of tunnel times from a large sample of readout measurements. In \autoref{fig:fid}a, we show two histograms, constructed with the same data-set, using different bin resolutions to distinguish between time scales. The histogram on the left in \autoref{fig:fid}a shows the fast-tunneling events. The tunnel rate for these events is faster than the detection bandwidth. We fit the histogram peak with a normal distribution, which appears to capture very well the readout statistic of the fast events, the detection of which is dominated by the rise time of the amplifier chain. The histogram on the right is fitted to an exponential decay corresponding to the slow tunnel-out rate of the singlet states, $\Gamma_{S\mathrm{,out}}$. We can then construct expressions for the error probabilities in detecting a $\ket{T}^z$ triplet ($P_{\rm error}^T$) or a singlet ($P_{\rm error}^S$), as a function of the detection time threshold $t_{\rm th}$:
\begin{align*}
P_{\rm error}^T(t_{\rm th}) &= 1 - \frac{1}{2}\left(1 + \mathrm{erf}\left(\frac{t_{\rm th} - \mu_d}{\sqrt{2\sigma_d^2}}\right)\right) \\
P_{\rm error}^S(t_{\rm th}) &= 1 - \exp\left(-\Gamma_{S\mathrm{,out}} t_{\rm th}\right)
\end{align*}

Here, $\mathrm{erf}(x) = 2/\sqrt{\pi} \int_0^x \exp(-t^2) dt$ is the error function. The error probabilities are plotted as a function of the detection threshold in \autoref{fig:fid}b.  We define the measurement fidelity in discriminating between fast ($\ket{T}^z$) and slow ($\ket{S}$) tunneling events as:
\begin{align*}
F_{\rm fs}(t_{\rm th}) = 1 - (P_{\rm error}^T(t_{\rm th}) + P_{\rm error}^S(t_{\rm th}))
\end{align*}

At the optimal value of $t_{\rm th} = 44$~$\mu$s we find a maximum fidelity $F_{\rm fs} = 95\%$.

The fidelity to discriminate between slow ($\ket{S}$) and no ($\ket{T}^{xy}$) tunneling events within the readout time window $t_R$ is simply given by:
\begin{align*}
F_{\rm sn} = P_{\rm error}^S(t_R)
\end{align*}

The fast ($\ket{T}^z$) events can be neglected, since $P_{\rm error}^T(t_R) \approx 0$ for large $t_R$. We find $F_{\rm sn} = 90\%$ at $t_R = 2$~ms (\autoref{fig:fid}b). Because of the very high signal-to-noise ratio in our time-resolved $I_{\rm SET}$ measurements, we have also neglected errors in the electrical detection of charge tunneling events in all the above analysis.

\end{document}